\documentclass[3p,times,procedia]{elsarticle}
\usepackage{nupha_ecrc}

\volume{00}

\firstpage{1}

\journalname{Nuclear Physics A}

\runauth{}

\jid{nupha}

\jnltitlelogo{Nuclear Physics A}

\usepackage{amssymb}

\usepackage[figuresright]{rotating}

\begin{document}

\begin{frontmatter}



\dochead{XXVIIth International Conference on Ultrarelativistic Nucleus-Nucleus Collisions\\ (Quark Matter 2018)}

\title{High-Temperature QCD: theory overview}


\author{Massimo D'Elia}

\address{Universit\`a di Pisa and INFN Sezione di Pisa,
Largo Pontecorvo 3, I-56127 Pisa, Italy}

\begin{abstract}
We review the recent progress achieved in the theoretical
investigation of Quantum Chromodynamics in the high temperature
regime, with a focus on results achieved by lattice QCD simulations.
The discussion covers the structure of the phase diagram and 
the properties of the strongly interacting medium at finite
$T$ and small baryon chemical potential.
\end{abstract}

\begin{keyword}
QCD Phase Diagram, Lattice QCD


\end{keyword}

\end{frontmatter}



\section{Introduction}
\label{sec:intro}

The theoretical study of QCD in the regime of high 
temperature and zero or small baryon chemical potential
has been a field of steady and continuous progress in the last few years.
Much progress has been obtained by means of lattice simulations, 
perturbation theory and 
effective model studies.
Our discussion 
is divided in three parts: 
in Section~\ref{finitettransition} we review recent studies
regarding the location and nature of the finite temperature 
transition as a function of the quark mass spectrum; 
Section~\ref{mediumprop} focuses on
various properties of the thermal medium; finally, 
in Section~\ref{finitemu}, we review recent results regarding 
the introduction of finite chemical potentials or other relevant extensions of
the QCD phase diagram.

\section{The finite temperature QCD transition}
\label{finitettransition}

Lattice QCD simulations have shown that, for
physical quark masses, the large
volume scaling 
around the temperature where
chiral symmetry restoration takes place
is consistent with the presence
of a rapid, analytic change of thermodynamical 
properties (what is usually called a crossover) in place of a 
real phase transition~\cite{Aoki:2006we}. The location
of the chiral pseudo-critical temperature $T_c$ has been 
well established by independent lattice studies
using different discretizations~\cite{Borsanyi:2010bp,Bazavov:2011nk};
a refinement has been presented at this conference,
indicating $T_c = 156.5 \pm  1.5$~MeV~\cite{Steinbrecher:2018phh}.

\begin{figure}[t!]
\centering
\includegraphics[scale=0.39]{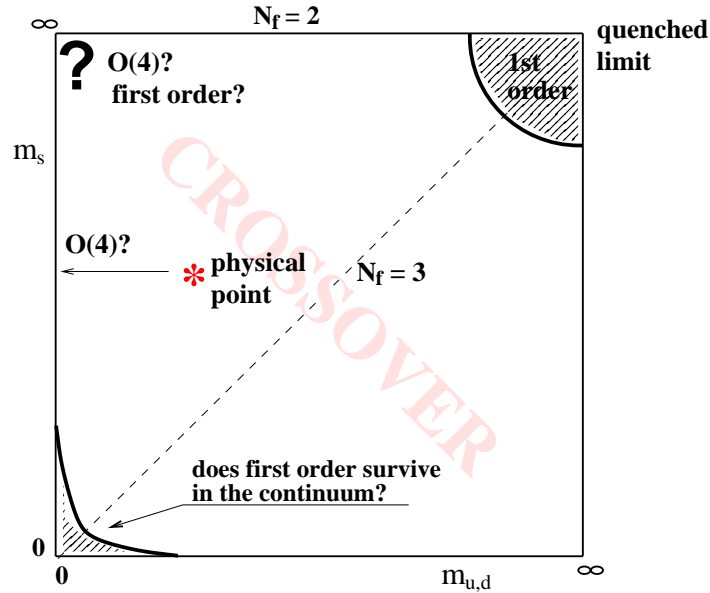}
\caption{Sketch of the nature of the finite temperature 
transition for $N_f = 2+1$ QCD as a function of the 
quark masses (Columbia plot).}
\label{fig:columbia}
\end{figure}

It is interesting, at least from a theoretical 
point of view, to determine how the nature of the 
transition changes as a function of the quark 
mass spectrum. This information is summarized,
for the theory with $N_f = 2+1$ flavours, in the so-called
Columbia plot, which is sketched in Fig.~\ref{fig:columbia}.
Most theoretical predictions are based
on universality arguments~\cite{Pisarski:1983ms},
through the analysis of an effective model 
sharing with QCD only the degrees of freedom involved
in the symmetry relevant to the transition. 
The standard argument
applies to cases where QCD symmetries are exact, hence 
to the upper-right corner (quenched limit with exact center symmetry)
and to the left border (chiral symmetry with massless up-down quarks): 
if the model does not show any fixed point, i.e.~any 
continuous transition with diverging correlation length,
then none is predicted for QCD as well and 
a first order transition is expected; on the contrary, 
if the model has a fixed point,
then QCD could still have a first order transition, but
a continuous transition in the same universality class 
of the effective model becomes the alternative possibility.
This kind of analysis extends partially to the rest of the Columbia plot,
since first order transitions are stable against
small explicit breakings of the relevant symmetry,
so a first order point implies a first order neighbourhood
around it, 
separated from the
crossover region by a second order line 
which is usually in the $3d$-Ising ($Z_2$) universality class,
because it delimits a region where two different phases emerge.

The quenched (pure gauge) limit is known to be first order,
while the analysis of effective chiral models predicts 
a first order transition
for $N_f \geq 3$ massless fermions (hence for the bottom-left corner)
and a possible second order (in the $O(4)$ universality class)
or a first order for $N_f = 2$ massless flavours;
in the latter case, the effective strength of the axial
$U_A(1)$ symmetry anomalous breaking around the transition 
could change the predicted universality 
class~\cite{Butti:2003nu,Basile:2005hw} or make
a second order transition unlikely.

However, checking these predictions
has revealed to be a hard task.
For instance, simulations of $N_f = 3$ and even $N_f = 4$ QCD  
have failed to determine a well defined 
continuum limit for the critical mass delimiting 
the chiral first order region~\cite{Endrodi:2007gc,deForcrand:2007rq,Jin:2014hea,Takeda:2016vfj,deForcrand:2017cgb}: the critical mass
could approach zero in the continuum, therefore
the question whether the first order
region in the bottom-left corner exists at all 
is still open.
One should consider,
in this respect, that while making reference to 
an effective chiral model seems perfectly reasonable,
there are various counter-examples of models in which 
the presence of gauge degrees of freedom can change
the situation substantially~\cite{Pelissetto:2017pxb,Pelissetto:2017sfd}.
On the other hand, the possible presence of a 
second order critical point
in the $O(4)$ universality class predicted 
in the chiral limit of $N_f = 2$ QCD (upper-left corner) has been
challenged by studies which suggest
a first order transition, at least on
coarse lattices~\cite{DElia:2005nmv,Bonati:2014kpa,Philipsen:2016hkv}.

An update has been reported at this conference~\cite{Ding:2018auz}
 regarding
the analysis of the chiral limit of 
the $N_f = 2+1$ theory, i.e.~leaving the strange quark 
mass at its physical value and approaching
the massless limit for the up and down quarks, for
QCD discretized via HISQ staggered quarks.
A first order transition is absent
down to pion masses as low as 80~MeV and
the scaling with the quark mass
of the chiral condensate and of its susceptibility, which 
is regulated by the critical index $\delta$, is consistent
with a chiral transition in the $O(4)$ (or $O(2)$, the residual
exact symmetry for staggered quarks) universality class,
$\delta \simeq 0.21$. That still does not exclude 
completely the possible presence of a $Z_2$ transition
at some finite but very low critical mass, the exponent
$\delta$ being practically indistinguishable in this case,
even if, also in view of the status in the $N_f = 3$ bottom-left
corner mentioned above, this is unlikely. Finally, 
a preliminary estimate has been provided for the critical 
temperature in the chiral and continuum limit,
$T_\chi = 138(5)$~MeV~\cite{Ding:2018auz}.

\section{Properties of the strongly interacting thermal medium}
\label{mediumprop}

Continuum extrapolated results for the equation of state
are available since a few years~\cite{Borsanyi:2013bia,Bazavov:2014pvz}
and have been obtained by lattice groups adopting
different staggered discretizations: the good 
agreement shows that systematic effects are well under control.
More recently, the computation has been extended 
to a region of significantly higher temperatures,
of the order of few GeVs,
both including~\cite{Borsanyi:2016ksw} 
and not including~\cite{Bazavov:2017dsy} charm quarks:
results reported at this conference
for $N_f=2+1$ QCD~\cite{Bazavov:2017dsy}
are in agreement with results 
from 3-loop HTL perturbation theory~\cite{Haque:2014rua}
and show, by comparison with Ref.~\cite{Borsanyi:2016ksw}, 
that charm contributions become appreciable 
when $T \gtrsim 500$ MeV.
A future challenge for lattice QCD is to bring computations
with discretizations other than staggered at the same 
level of accuracy; promising approaches
to improve on Wilson fermions are on the way,
based on the imaginary moving 
frame approach~\cite{Giusti:2010bb,Giusti:2014ila,DallaBrida:2017sxr}, 
on the gradient flow~\cite{Suzuki:2013gza,Asakawa:2013laa,Kitazawa:2016dsl},
on non-equilibrium methods~\cite{Caselle:2018kap} and
on twisted mass discretizations~\cite{Burger:2015xda}.

The analysis of the equation of state is not exhaustive
of the many interesting changes occurring in the thermal medium
across $T_c$. The fate of the confining properties
is another issue of obvious theoretical and
phenomenological interest. Deconfinement is clearly 
signalled by the suppression of
color flux tubes across $T_c$~\cite{Cea:2015wjd,Bicudo:2017uyy,Cea:2017bsa},
even if structures similar to flux tubes still seem to form 
between heavy $\bar Q Q$ pairs
for temperatures up 
to $\sim 1.5\, T_c$~\cite{Cea:2017bsa}, and have been related
to the possible presence of chromomagnetic charges populating
the Quark-Gluon Plasma (QGP)~\cite{Shuryak:2018ytg}. 
Precise determinations of the heavy $\bar Q Q$ potential 
at higher temperatures have been reported 
recently~\cite{Bazavov:2018wmo} (for $T$ up to $\sim 6$~GeV), showing that 
the potential is perturbative at short distances
and screened for distances $r$ such that $r\, T \gtrsim 0.3$;
results on the dependence of color screening on the baryon
density have also been reported at this conference~\cite{Andreoli:2017zie}.
The investigation of $\bar Q Q$ interactions 
is essential
to clarify the fate and suppression of heavy quark bound states
within the thermal medium: many direct studies of such states
have been reported recently~\cite{Ding:2018uhl,Kelly:2018hsi}, 
and progress has been made also in the determination
of a realistic in-medium potential~\cite{Kim:2018yhk},
which is the correct quantity to be used in place 
of static $\bar Q Q$ free energies.

It is also interesting to ask what is the effect 
on the properties of the QGP of chiral 
symmetry restoration: this is visible, for instance,
from the recovered degeneracy of various meson
correlators. More recently, this has been
clearly observed also in the baryon sector, 
where different parity doublets become degenerate as 
$T$ approaches $T_c$ from below~\cite{Aarts:2017rrl,Aarts:2017gmj}.
Also the axial $U_A(1)$ symmetry, even if always explicitly 
broken by the presence of the axial anomaly, is expected
to undergo an effective restoration, because of the suppression
of gauge configurations with non-zero winding number, and various
studies show that this happens for 
$T \gtrsim 200$~MeV~\cite{Tomiya:2016jwr,Fukaya:2017wfq,Bhattacharya:2014ara,Chiu:2013wwa}.
It is also interesting to stress the 
unexpected degeneracy which has been recently observed
in meson correlators for
$T \gtrsim 2\, T_c$~\cite{Rohrhofer:2018pey}
and which could reveal the presence of an extended symmetry
group for the QGP, larger than what expected based just on chiral symmetry:
that could shed light on the relevance of chromomagnetic and chromoelectric 
interactions in that temperature regime.

Concerning the transport properties of the thermal medium,
their study by non-perturbative lattice simulations is 
notoriously 
difficult, because they are related
to non-equilibrium properties.
In principle, Euclidean correlators give direct access 
to the relevant spectral functions and the transport 
coefficients can be obtained by solving 
appropriate integral equations~\cite{Meyer:2011gj},
however that requires to have an extreme accuracy on 
the correlators, which
is presently achievable only in the 
quenched theory: results for the heavy quark diffusion coefficient
have been reported three years ago~\cite{Francis:2015daa},
while updated results for the bulk and shear viscosity 
have been discussed at this 
conference~\cite{Astrakhantsev:2017nrs,Astrakhantsev:2018oue}.
Unfortunately, such high precision is not yet 
achievable in full QCD and
success is limited to a few cases,
like the computation of the electric 
conductivity~\cite{Amato:2013naa,Aarts:2014nba};
hopes for the 
future are linked to the development
of efficient  multilevel updating 
schemes~\cite{Ce:2016idq}.
Substantial progress has been reported 
at this conference on the computation of 
transport coefficients by perturbation
theory~\cite{Ghiglieri:2018dib}, or by model 
studies which take lattice data as an input~\cite{Liu:2018ons}.

\section{Finite temperature QCD in the presence of external sources}
\label{finitemu}

When one tries to extend the exploration of the 
QCD phase diagram to the finite density case,
one has to face the well known sign problem. One can investigate
systematically only special cases, 
like that of matter with zero overall baryonic charge
but with a finite density of isospin charge (e.g., a pion gas),
for which recent results can be found in 
Ref.~\cite{Brandt:2017oyy}. Instead, when one tries to switch
on a finite baryonic chemical potential, the path integral
measure becomes complex, thus hindering the exploitation
of standard Monte-Carlo sampling techniques.
While no definite solution has still been found, some partial solutions
are known to work well in a regime of small values of 
$\mu_B/T$, namely Taylor expansion~\cite{tayl1,tayl2,tayl3,tayl4},
analytic continuation
from an imaginary chemical potential~\cite{imag1,imag2,imag3}.

A quantity for which these approximate techniques are well suited 
is the curvature of the pseudocritical line in 
the $T - \mu_B$ plane separating confined phase from the
QGP phase, which is defined by
\begin{equation}
\frac{T_c(\mu_B)}{T_c}=1-\kappa \left(\frac{\mu_B}{T_c}\right)^2\, +\, 
O(\mu_B^4)\, .
\label{curvature}
\end{equation}
{The curvature $\kappa$ can be determined by following 
{explicitly} how  $T_c$ moves at imaginary $\mu_B$ and then continuing 
to real $\mu_B$, or by finding $d T_c / d \mu_B^2$ {implicitly} in terms
of derivatives computed by lattice simulations 
at $\mu_B = 0$ (Taylor expansion). While these techniques have 
been applied since long, it is useful to mention some
recent determinations, obtained with discretizations of QCD
which are close to the physical point~\cite{Kaczmarek:2011zz,Endrodi:2011gv,Cea:2014xva,Bonati:2014rfa,Bonati:2015bha,Bellwied:2015rza,Cea:2015cya,Hegde:2015tbn,Steinbrecher:2018phh,Bonati:2018nut},
and are summarized in Fig.~\ref{fig:kappa}; two of those
determinations have been presented 
at this conference~\cite{Steinbrecher:2018phh,Bonati:2018nut,Bonati:2018wdn}
and represent an important confirmation that, after systematic effects
have been properly taken into account, 
Taylor expansion and analytic continuation lead to consistent and 
reliable results.
The various determinations have been obtained for a variety of 
choices of the strange quark chemical potential: 
$\mu_s =0$~\cite{Kaczmarek:2011zz,Endrodi:2011gv,Bonati:2014rfa,Bonati:2015bha,Bonati:2018nut}, 
$\mu_s = \mu_l$~\cite{Cea:2014xva,Cea:2015cya}
or tuned so as to guarantee strangeness 
neutrality~\cite{Bellwied:2015rza,Steinbrecher:2018phh}
as in heavy ion collisions; as a matter of fact, one finds that 
$\mu_s$ has a negligible effect
on $\kappa$~\cite{Bonati:2014rfa,Steinbrecher:2018phh}.
In the figure we report also an average (vertical dashed band)
of the five most recent 
determinations obtained after continuum extrapolation (filled dots),
this average is $\kappa = 0.014(2)$ and is quite stable 
when excluding one or more of the mentioned determinations; we have been
conservative with the error estimate, since the errors
of the various determinations include also systematic contributions,
which are tipically different for the different methods.

\begin{figure}[t!]
\centering
\includegraphics[scale=0.4]{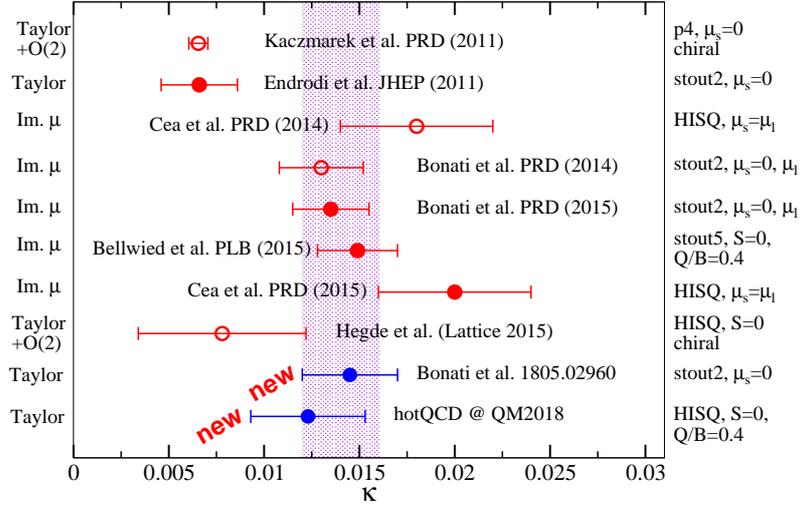}
\caption{Summary of recent determinations of the curvature
of the pseudo-critical line, 
defined in Eq.~(\ref{curvature}),
obtained at or close to the physical
point. For each determination, 
the method adopted to extract $\kappa$ is indicated
on the left, while the adopted discretization and the setup of 
chemical potentials are indicated on the right. Chiral and $O(2)$ 
refer to determinations obtained in the chiral limit assuming
a second order critical scaling. Finally, 
filled circles indicate determinations
obtained after continuum extrapolation. The vertical dashed band
is an average of continuum extrapolated results obtained after 2015.}
\label{fig:kappa}
\end{figure}

Other quantities which can be investigated systematically even 
with the available numerical approaches are the
so-called generalized
susceptibilities, i.e.~higher order derivatives 
of the partition function with respect
to the quark chemical potentials:
\begin{equation}
\label{free_energy_expansion}
\frac{P}{T^4}(\mu_u,\mu_d,\mu_s) = 
\frac{P}{T^4}(0,0,0) 
+
\sum_{i+j+k=even}\hspace{-0.3cm}\frac{\chi_{ijk}(T)}{i!j!k!}\hat{\mu}_{u}^{(i)}\hat{\mu}_{d}^{(j)}\hat{\mu}_{s}^{(k)} \ \  ; \ \ \ \ 
\chi_{ijk}(T) =
\frac{1}{VT^{4}}\frac{\partial^{(i+j+k)}F(T,\mu)}{\partial{\hat{\mu}_{u}}^{(i)}\partial{\hat{\mu}_{d}}^{(j)}\partial{\hat{\mu}_{s}}^{(k)}}\bigg|_{\mu_u
  = \mu_d = \mu_s = 0} \, .
\end{equation}
As for other quantities related to the dependence of the free energy
density on the chemical potentials, the 
coefficients $\chi_{ijk}(T)$ 
can be determined as Taylor coefficients computed at zero chemical 
potential~\cite{tayl1,tayl2,tayl3,tayl4,tayl5}, or exploiting
numerical simulations at imaginary $\mu$ and then analytic 
continuation~\cite{DElia:2004ani,DElia:2009pdy,Takaishi:2010kc,Gunther:2016vcp,DElia:2016jqh}. Using the latter method one can reach up to the 
eighth order in the expansion~\cite{DElia:2016jqh}: 
new results along this line have been presented 
at this conference~\cite{Borsanyi:2018grb}.

Generalized susceptibilities are important for various phenomenological
and theoretical reasons.
For instance, since they are directly 
related to fluctuations of conserved charges,
they can be used to compare the QCD thermal medium
with the experimental output from heavy ion collisions
and obtain a model-independent 
determination 
of the freeze-out line~\cite{Bazavov:2012vg,Bazavov:2015zja,Borsanyi:2013hza};
an update of research along this line 
has been reported at this conference 
(posters by C.~Ratti and C.~Schmidt). They are
also extremely useful as an input or benchmark for a variety
of model approaches to the strongly interacting 
medium~\cite{Almasi:2018lok,Liu:2016ysz,Liu:2017qah,Vovchenko:2017gkg,Huovinen:2017ogf,Parotto:2018pwx}.
}

Besides that, generalized susceptibilities can be used 
to investigate the properties of convergence of 
the Taylor series: this is useful since the possible presence of a 
critical point in the QCD phase diagram is signalled by 
a finite radius of convergence. The radius can be estimated
directly from the 
series expansion of the free energy density, or from those of
its derivatives:
\begin{equation}
\rho_{n,m}^{f}= \left(
\frac{\chi_{n}^{B}/n!}{\chi_{m}^{B}/m!}\right)^{\frac{1}{(m-n)}} \ \ ; \ \ \ \ 
\rho_{n,m}^{\chi}= \left(
\frac{\chi_{n}^{B}/(n-2)!}{\chi_{m}^{B}/(m-2)!}\right)^{\frac{1}{(m-n)}}
\, .
\end{equation}
In presence of a finite radius of convergence both
definitions should converge to a common value
as $n,m~\to~\infty$; a correct estimate of 
the radius is thus linked to the possibility 
of accessing the asymptotic behaviour of the series: however, what
``asymptotic'' means is a priori not known, and 
estimates based on just the first 3-4 terms might be affected 
by unknown systematics.
 Various studies have exploited this 
method~\cite{Allton:2003vx,tayl4,Gavai:2008zr}, including 
some studies appeared recently~\cite{Datta:2016ukp,DElia:2016jqh,tayl5};
new results have been reported 
at this conference as well~\cite{Fodor:2018wul}. 
The overall outcome (see Ref.~\cite{Fodor:2018wul} for
more details) is that in some studies no convergence 
of the radius estimate is observed and
that, even when this is observed by looking at just 
the first few ratios,
one should be careful, since it could be fake.
One should take into account the possibility that,
even in the presence of a critical point, lattice data could
not be sensitive to it, or we should be able access much higher 
orders in the series expansion, something which is currently unfeasible. 
In this respect, much insight
could come from models which take lattice data as an input~\cite{Vovchenko:2018zgt, Parotto:2018gic}.
The fact that no critical point is likely to be found close
to $\mu_B = 0$ is also supported by the apparent lack of strong
$\mu$-dependence of various 
susceptibilities (see Refs.~\cite{Steinbrecher:2018phh,Bonati:2015bha}
and talk by S.~Borsanyi at this conference).

There are other extensions of the QCD phase diagram which are of 
interest for heavy ion collisions. One of them is 
the introduction of a magnetic background field,
which is relevant to at least the early stages of 
the collisions. Many things are well known,
like the fact that the QGP behaves as a paramagnetic medium,
with a magnetic susceptibility which strongly increases
with the temperature~\cite{Bonati:2013lca,Levkova:2013qda,Bonati:2013vba,Bali:2013owa,Bali:2014kia}. The pseudo-critical temperature decreases
as a function of $B$~\cite{Bali:2011qj}, and 
recent results suggest that this behaviour
is qualitatively independent of the quark mass 
spectrum~\cite{DElia:2018xwo}. 
Another interesting extension, of obvious interest to peripherical 
collisions, is the study of a rotating thermal medium;
in this case the sign problem is back again, however 
preliminary studies exist which should be further pursued 
in the future~\cite{Yamamoto:2013zwa}.






\begin{thebibliography}{100}
\expandafter\ifx\csname url\endcsname\relax
  \def\url#1{\texttt{#1}}\fi
\expandafter\ifx\csname urlprefix\endcsname\relax\def\urlprefix{URL }\fi
\expandafter\ifx\csname href\endcsname\relax
  \def\href#1#2{#2} \def\path#1{#1}\fi

\bibitem{Aoki:2006we}
Y.~Aoki, G.~Endrodi, Z.~Fodor, S.~D. Katz, K.~K. Szabo, Nature 443 (2006)
  675--678,
\newblock \href {http://arxiv.org/abs/hep-lat/0611014}
  {\path{arXiv:hep-lat/0611014}}.

\bibitem{Borsanyi:2010bp}
S.~Borsanyi, Z.~Fodor, C.~Hoelbling, S.~D. Katz, S.~Krieg, C.~Ratti, K.~K.
  Szabo, JHEP 09 (2010) 073,
\newblock \href {http://arxiv.org/abs/1005.3508} {\path{arXiv:1005.3508}}.

\bibitem{Bazavov:2011nk}
A.~Bazavov, et~al., Phys. Rev. D85 (2012) 054503,
\newblock \href {http://arxiv.org/abs/1111.1710} {\path{arXiv:1111.1710}}.

\bibitem{Steinbrecher:2018phh}
P.~Steinbrecher, \href {http://arxiv.org/abs/1807.05607}
  {\path{arXiv:1807.05607}}.

\bibitem{Pisarski:1983ms}
R.~D. Pisarski, F.~Wilczek, Phys. Rev. D29 (1984) 338--341.

\bibitem{Butti:2003nu}
A.~Butti, A.~Pelissetto, E.~Vicari, JHEP 08 (2003) 029, 
\newblock \href {http://arxiv.org/abs/hep-ph/0307036}
  {\path{arXiv:hep-ph/0307036}}.

\bibitem{Basile:2005hw}
F.~Basile, A.~Pelissetto, E.~Vicari, PoS LAT2005 (2006) 199, 
\newblock \href {http://arxiv.org/abs/hep-lat/0509018}
  {\path{arXiv:hep-lat/0509018}}.

\bibitem{Endrodi:2007gc}
G.~Endrodi, Z.~Fodor, S.~D. Katz, K.~K. Szabo, PoS LATTICE2007 (2007) 182,
\newblock \href {http://arxiv.org/abs/0710.0998} {\path{arXiv:0710.0998}}.

\bibitem{deForcrand:2007rq}
P.~de~Forcrand, S.~Kim, O.~Philipsen, PoS LATTICE2007 (2007) 178,
\newblock \href {http://arxiv.org/abs/0711.0262} {\path{arXiv:0711.0262}}.

\bibitem{Jin:2014hea}
X.-Y. Jin, Y.~Kuramashi, Y.~Nakamura, S.~Takeda, A.~Ukawa, Phys. Rev. D91~(1)
  (2015) 014508,
\newblock \href {http://arxiv.org/abs/1411.7461} {\path{arXiv:1411.7461}}.

\bibitem{Takeda:2016vfj}
S.~Takeda, X.-Y. Jin, Y.~Kuramashi, Y.~Nakamura, A.~Ukawa, PoS LATTICE2016
  (2017) 384,
\newblock \href {http://arxiv.org/abs/1612.05371} {\path{arXiv:1612.05371}}.

\bibitem{deForcrand:2017cgb}
P.~de~Forcrand, M.~D'Elia, PoS LATTICE2016 (2017) 081,
\newblock \href {http://arxiv.org/abs/1702.00330} {\path{arXiv:1702.00330}}.

\bibitem{Pelissetto:2017pxb}
A.~Pelissetto, A.~Tripodo, E.~Vicari, Phys. Rev. E97~(1) (2018) 012123,
\newblock \href {http://arxiv.org/abs/1711.04567} {\path{arXiv:1711.04567}}.

\bibitem{Pelissetto:2017sfd}
A.~Pelissetto, A.~Tripodo, E.~Vicari, Phys. Rev. D96~(3) (2017) 034505,
\newblock \href {http://arxiv.org/abs/1706.04365} {\path{arXiv:1706.04365}}.

\bibitem{DElia:2005nmv}
M.~D'Elia, A.~Di~Giacomo, C.~Pica, Phys. Rev. D72 (2005) 114510,
\newblock \href {http://arxiv.org/abs/hep-lat/0503030}
  {\path{arXiv:hep-lat/0503030}}.

\bibitem{Bonati:2014kpa}
C.~Bonati, P.~de~Forcrand, M.~D'Elia, O.~Philipsen, F.~Sanfilippo, Phys. Rev.
  D90~(7) (2014) 074030,
\newblock \href {http://arxiv.org/abs/1408.5086} {\path{arXiv:1408.5086}}.

\bibitem{Philipsen:2016hkv}
O.~Philipsen, C.~Pinke, Phys. Rev. D93~(11) (2016) 114507,
\newblock \href {http://arxiv.org/abs/1602.06129} {\path{arXiv:1602.06129}}.

\bibitem{Ding:2018auz}
H.~T. Ding, P.~Hegde, F.~Karsch, A.~Lahiri, S.~T. Li, S.~Mukherjee,
  P.~Petreczky,
\newblock \href {http://arxiv.org/abs/1807.05727} {\path{arXiv:1807.05727}}.

\bibitem{Borsanyi:2013bia}
S.~Borsanyi, Z.~Fodor, C.~Hoelbling, S.~D. Katz, S.~Krieg, K.~K. Szabo, Phys.
  Lett. B730 (2014) 99--104,
\newblock \href {http://arxiv.org/abs/1309.5258} {\path{arXiv:1309.5258}}.

\bibitem{Bazavov:2014pvz}
A.~Bazavov, et~al., Phys. Rev. D90 (2014) 094503,
\newblock \href {http://arxiv.org/abs/1407.6387} {\path{arXiv:1407.6387}}.

\bibitem{Borsanyi:2016ksw}
S.~Borsanyi, et~al., Nature 539~(7627) (2016) 69--71,
\newblock \href {http://arxiv.org/abs/1606.07494} {\path{arXiv:1606.07494}}.

\bibitem{Bazavov:2017dsy}
A.~Bazavov, P.~Petreczky, J.~H. Weber, Phys. Rev. D97~(1) (2018) 014510,
\newblock \href {http://arxiv.org/abs/1710.05024} {\path{arXiv:1710.05024}}.

\bibitem{Haque:2014rua}
N.~Haque, A.~Bandyopadhyay, J.~O. Andersen, M.~G. Mustafa, M.~Strickland,
  N.~Su, JHEP 05 (2014) 027,
\newblock \href {http://arxiv.org/abs/1402.6907} {\path{arXiv:1402.6907}}.

\bibitem{Giusti:2010bb}
L.~Giusti, H.~B. Meyer, Phys. Rev. Lett. 106 (2011) 131601,
\newblock \href {http://arxiv.org/abs/1011.2727} {\path{arXiv:1011.2727}}.

\bibitem{Giusti:2014ila}
L.~Giusti, M.~Pepe, Phys. Rev. Lett. 113 (2014) 031601,
\newblock \href {http://arxiv.org/abs/1403.0360} {\path{arXiv:1403.0360}}.

\bibitem{DallaBrida:2017sxr}
M.~Dalla~Brida, L.~Giusti, M.~Pepe, EPJ Web Conf. 175 (2018) 14012,
\newblock \href {http://arxiv.org/abs/1710.09219} {\path{arXiv:1710.09219}}.

\bibitem{Suzuki:2013gza}
H.~Suzuki, PTEP 2013 (2013) 083B03, [Erratum: PTEP2015,079201(2015)],
\newblock \href {http://arxiv.org/abs/1304.0533} {\path{arXiv:1304.0533}}.

\bibitem{Asakawa:2013laa}
M.~Asakawa, T.~Hatsuda, E.~Itou, M.~Kitazawa, H.~Suzuki, Phys. Rev. D90~(1)
  (2014) 011501, [Erratum: Phys. Rev.D92,no.5,059902(2015)],
\newblock \href {http://arxiv.org/abs/1312.7492} {\path{arXiv:1312.7492}}.

\bibitem{Kitazawa:2016dsl}
M.~Kitazawa, T.~Iritani, M.~Asakawa, T.~Hatsuda, H.~Suzuki, Phys. Rev. D94~(11)
  (2016) 114512,
\newblock \href {http://arxiv.org/abs/1610.07810} {\path{arXiv:1610.07810}}.

\bibitem{Caselle:2018kap}
M.~Caselle, A.~Nada, M.~Panero, \href {http://arxiv.org/abs/1801.03110}
  {\path{arXiv:1801.03110}}.

\bibitem{Burger:2015xda}
F.~Burger, E.-M. Ilgenfritz, M.~P. Lombardo, M.~Muller-Preussker, A.~Trunin, J.
  Phys. Conf. Ser. 668~(1) (2016) 012092,
\newblock \href {http://arxiv.org/abs/1510.02262} {\path{arXiv:1510.02262}}.

\bibitem{Cea:2015wjd}
P.~Cea, L.~Cosmai, F.~Cuteri, A.~Papa, JHEP 06 (2016) 033,
\newblock \href {http://arxiv.org/abs/1511.01783} {\path{arXiv:1511.01783}}.

\bibitem{Bicudo:2017uyy}
P.~Bicudo, N.~Cardoso, M.~Cardoso, \href {http://arxiv.org/abs/1702.03454}
  {\path{arXiv:1702.03454}}.

\bibitem{Cea:2017bsa}
P.~Cea, L.~Cosmai, F.~Cuteri, A.~Papa, EPJ Web Conf. 175 (2018) 12006,
\newblock \href {http://arxiv.org/abs/1710.01963} {\path{arXiv:1710.01963}}.

\bibitem{Shuryak:2018ytg}
E.~Shuryak, \href {http://arxiv.org/abs/1806.10487} {\path{arXiv:1806.10487}}.

\bibitem{Bazavov:2018wmo}
A.~Bazavov, N.~Brambilla, P.~Petreczky, A.~Vairo, J.~H. Weber, \href
  {http://arxiv.org/abs/1804.10600} {\path{arXiv:1804.10600}}.

\bibitem{Andreoli:2017zie}
M.~Andreoli, C.~Bonati, M.~D'Elia, M.~Mesiti, F.~Negro, A.~Rucci,
  F.~Sanfilippo, Phys. Rev. D97~(5) (2018) 054515,
\newblock \href {http://arxiv.org/abs/1712.09996} {\path{arXiv:1712.09996}}.

\bibitem{Ding:2018uhl}
H.-T. Ding, O.~Kaczmarek, A.-L. Kruse, R.~Larsen, L.~Mazur, S.~Mukherjee,
  H.~Ohno, H.~Sandmeyer, H.-T. Shu,
\newblock \href {http://arxiv.org/abs/1807.06315} {\path{arXiv:1807.06315}}.

\bibitem{Kelly:2018hsi}
A.~Kelly, A.~Rothkopf, J.-I. Skullerud, Phys. Rev. D97~(11) (2018) 114509,
\newblock \href {http://arxiv.org/abs/1802.00667} {\path{arXiv:1802.00667}}.

\bibitem{Kim:2018yhk}
S.~Kim, P.~Petreczky, A.~Rothkopf, \href {http://arxiv.org/abs/1808.08781}
  {\path{arXiv:1808.08781}}.

\bibitem{Aarts:2017rrl}
G.~Aarts, C.~Allton, D.~De~Boni, S.~Hands, B.~Jager, C.~Praki, J.-I. Skullerud,
  JHEP 06 (2017) 034,
\newblock \href {http://arxiv.org/abs/1703.09246} {\path{arXiv:1703.09246}}.

\bibitem{Aarts:2017gmj}
G.~Aarts, C.~Allton, D.~De~Boni, S.~Hands, B.~Jager, C.~Praki, J.-I. Skullerud,
  EPJ Web Conf. 175 (2018) 07016, 
\newblock \href {http://arxiv.org/abs/1710.08294} {\path{arXiv:1710.08294}}.

\bibitem{Tomiya:2016jwr}
A.~Tomiya, G.~Cossu, S.~Aoki, H.~Fukaya, S.~Hashimoto, T.~Kaneko, J.~Noaki,
  Phys. Rev. D96~(3) (2017) 034509, [Addendum: Phys.
  Rev.D96,no.7,079902(2017)], 
\newblock \href {http://arxiv.org/abs/1612.01908} {\path{arXiv:1612.01908}}.

\bibitem{Fukaya:2017wfq}
H.~Fukaya, EPJ Web Conf. 175 (2018) 01012,
\newblock \href {http://arxiv.org/abs/1712.05536} {\path{arXiv:1712.05536}}.

\bibitem{Bhattacharya:2014ara}
T.~Bhattacharya, et~al., Phys. Rev. Lett. 113~(8) (2014) 082001, 
\newblock \href {http://arxiv.org/abs/1402.5175} {\path{arXiv:1402.5175}}.

\bibitem{Chiu:2013wwa}
T.-W. Chiu, W.-P. Chen, Y.-C. Chen, H.-Y. Chou, T.-H. Hsieh, PoS LATTICE2013
  (2014) 165, 
\newblock \href {http://arxiv.org/abs/1311.6220} {\path{arXiv:1311.6220}}.

\bibitem{Rohrhofer:2018pey}
C.~Rohrhofer, Y.~Aoki, G.~Cossu, L.~{\relax Ya}. Glozman, S.~Hashimoto,
  S.~Prelovsek, 2018, 
\newblock \href {http://arxiv.org/abs/1809.00244} {\path{arXiv:1809.00244}}.

\bibitem{Meyer:2011gj}
H.~B. Meyer, Eur. Phys. J. A47 (2011) 86, 
\newblock \href {http://arxiv.org/abs/1104.3708} {\path{arXiv:1104.3708}}.

\bibitem{Francis:2015daa}
A.~Francis, O.~Kaczmarek, M.~Laine, T.~Neuhaus, H.~Ohno, Phys. Rev. D92~(11)
  (2015) 116003, 
\newblock \href {http://arxiv.org/abs/1508.04543} {\path{arXiv:1508.04543}}.

\bibitem{Astrakhantsev:2017nrs}
N.~Astrakhantsev, V.~Braguta, A.~Kotov, JHEP 04 (2017) 101, 
\newblock \href {http://arxiv.org/abs/1701.02266} {\path{arXiv:1701.02266}}.

\bibitem{Astrakhantsev:2018oue}
N.~{\relax Yu}. Astrakhantsev, V.~V. Braguta, A.~{\relax Yu}. Kotov, \href
  {http://arxiv.org/abs/1804.02382} {\path{arXiv:1804.02382}}.

\bibitem{Amato:2013naa}
A.~Amato, G.~Aarts, C.~Allton, P.~Giudice, S.~Hands, J.-I. Skullerud, Phys.
  Rev. Lett. 111~(17) (2013) 172001, 
\newblock \href {http://arxiv.org/abs/1307.6763} {\path{arXiv:1307.6763}}.

\bibitem{Aarts:2014nba}
G.~Aarts, C.~Allton, A.~Amato, P.~Giudice, S.~Hands, J.-I. Skullerud, JHEP 02
  (2015) 186, 
\newblock \href {http://arxiv.org/abs/1412.6411} {\path{arXiv:1412.6411}}.

\bibitem{Ce:2016idq}
M.~C\`e, L.~Giusti, S.~Schaefer, Phys. Rev. D93~(9) (2016) 094507, 
\newblock \href {http://arxiv.org/abs/1601.04587} {\path{arXiv:1601.04587}}.

\bibitem{Ghiglieri:2018dib}
J.~Ghiglieri, G.~D. Moore, D.~Teaney, JHEP 03 (2018) 179, 
\newblock \href {http://arxiv.org/abs/1802.09535} {\path{arXiv:1802.09535}}.

\bibitem{Liu:2018ons}
S.~Y.~F. Liu, R.~Rapp, 
\newblock \href {http://arxiv.org/abs/1807.06739} {\path{arXiv:1807.06739}}.

\bibitem{Brandt:2017oyy}
B.~B. Brandt, G.~Endrodi, S.~Schmalzbauer, Phys. Rev. D97~(5) (2018) 054514,
\newblock \href {http://arxiv.org/abs/1712.08190} {\path{arXiv:1712.08190}}.

\bibitem{tayl1}
C.~R. Allton, S.~Ejiri, S.~J. Hands, O.~Kaczmarek, F.~Karsch, E.~Laermann,
  C.~Schmidt, L.~Scorzato, Phys. Rev. D66 (2002) 074507, 
\newblock \href {http://arxiv.org/abs/hep-lat/0204010}
  {\path{arXiv:hep-lat/0204010}}.

\bibitem{tayl2}
C.~R. Allton, M.~Doring, S.~Ejiri, S.~J. Hands, O.~Kaczmarek, F.~Karsch,
  E.~Laermann, K.~Redlich, Phys. Rev. D71 (2005) 054508, 
\newblock \href {http://arxiv.org/abs/hep-lat/0501030}
  {\path{arXiv:hep-lat/0501030}}.

\bibitem{tayl3}
R.~V. Gavai, S.~Gupta, Phys. Rev. D68 (2003) 034506, 
\newblock \href {http://arxiv.org/abs/hep-lat/0303013}
  {\path{arXiv:hep-lat/0303013}}.

\bibitem{tayl4}
R.~V. Gavai, S.~Gupta, Phys. Rev. D71 (2005) 114014, 
\newblock \href {http://arxiv.org/abs/hep-lat/0412035}
  {\path{arXiv:hep-lat/0412035}}.

\bibitem{imag1}
M.-P. Lombardo, Nucl. Phys. Proc. Suppl. 83 (2000) 375--377, 
\newblock \href {http://arxiv.org/abs/hep-lat/9908006}
  {\path{arXiv:hep-lat/9908006}}.

\bibitem{imag2}
P.~de~Forcrand, O.~Philipsen, Nucl. Phys. B642 (2002) 290--306, 
\newblock \href {http://arxiv.org/abs/hep-lat/0205016}
  {\path{arXiv:hep-lat/0205016}}.

\bibitem{imag3}
M.~D'Elia, M.-P. Lombardo, Phys. Rev. D67 (2003) 014505, 
\newblock \href {http://arxiv.org/abs/hep-lat/0209146}
  {\path{arXiv:hep-lat/0209146}}.

\bibitem{Kaczmarek:2011zz}
O.~Kaczmarek, F.~Karsch, E.~Laermann, C.~Miao, S.~Mukherjee, P.~Petreczky,
  C.~Schmidt, W.~Soeldner, W.~Unger, Phys. Rev. D83 (2011) 014504, 
\newblock \href {http://arxiv.org/abs/1011.3130} {\path{arXiv:1011.3130}}.

\bibitem{Endrodi:2011gv}
G.~Endrodi, Z.~Fodor, S.~D. Katz, K.~K. Szabo, JHEP 04 (2011) 001, 
\newblock \href {http://arxiv.org/abs/1102.1356} {\path{arXiv:1102.1356}}.

\bibitem{Cea:2014xva}
P.~Cea, L.~Cosmai, A.~Papa, Phys. Rev. D89~(7) (2014) 074512, 
\newblock \href {http://arxiv.org/abs/1403.0821} {\path{arXiv:1403.0821}}.

\bibitem{Bonati:2014rfa}
C.~Bonati, M.~D'Elia, M.~Mariti, M.~Mesiti, F.~Negro, F.~Sanfilippo, Phys. Rev.
  D90~(11) (2014) 114025, 
\newblock \href {http://arxiv.org/abs/1410.5758} {\path{arXiv:1410.5758}}.

\bibitem{Bonati:2015bha}
C.~Bonati, M.~D'Elia, M.~Mariti, M.~Mesiti, F.~Negro, F.~Sanfilippo, Phys. Rev.
  D92~(5) (2015) 054503, 
\newblock \href {http://arxiv.org/abs/1507.03571} {\path{arXiv:1507.03571}}.

\bibitem{Bellwied:2015rza}
R.~Bellwied, S.~Borsanyi, Z.~Fodor, J.~Günther, S.~D. Katz, C.~Ratti, K.~K.
  Szabo, Phys. Lett. B751 (2015) 559--564, 
\newblock \href {http://arxiv.org/abs/1507.07510} {\path{arXiv:1507.07510}}.

\bibitem{Cea:2015cya}
P.~Cea, L.~Cosmai, A.~Papa, Phys. Rev. D93~(1) (2016) 014507, 
\newblock \href {http://arxiv.org/abs/1508.07599} {\path{arXiv:1508.07599}}.

\bibitem{Hegde:2015tbn}
P.~Hegde, H.-T. Ding, PoS LATTICE2015 (2016) 141, 
\newblock \href {http://arxiv.org/abs/1511.03378} {\path{arXiv:1511.03378}}.

\bibitem{Bonati:2018nut}
C.~Bonati, M.~D'Elia, F.~Negro, F.~Sanfilippo, K.~Zambello, Phys. Rev. D98~(5)
  (2018) 054510, 
\newblock \href {http://arxiv.org/abs/1805.02960} {\path{arXiv:1805.02960}}.

\bibitem{Bonati:2018wdn}
C.~Bonati, M.~D'Elia, F.~Negro, F.~Sanfilippo, K.~Zambello, \href
  {http://arxiv.org/abs/1807.10026} {\path{arXiv:1807.10026}}.

\bibitem{tayl5}
A.~Bazavov, et~al., Phys. Rev. D95~(5) (2017) 054504, 
\newblock \href {http://arxiv.org/abs/1701.04325} {\path{arXiv:1701.04325}}.

\bibitem{DElia:2004ani}
M.~D'Elia, M.~P. Lombardo, Phys. Rev. D70 (2004) 074509, 
\newblock \href {http://arxiv.org/abs/hep-lat/0406012}
  {\path{arXiv:hep-lat/0406012}}.

\bibitem{DElia:2009pdy}
M.~D'Elia, F.~Sanfilippo, Phys. Rev. D80 (2009) 014502, 
\newblock \href {http://arxiv.org/abs/0904.1400} {\path{arXiv:0904.1400}}.

\bibitem{Takaishi:2010kc}
T.~Takaishi, P.~de~Forcrand, A.~Nakamura, PoS LAT2009 (2009) 198,
\newblock \href {http://arxiv.org/abs/1002.0890} {\path{arXiv:1002.0890}}.

\bibitem{Gunther:2016vcp}
J.~N. Guenther, R.~Bellwied, S.~Borsanyi, Z.~Fodor, S.~D. Katz, A.~Pasztor,
  C.~Ratti, K.~K. Szabo, Nucl. Phys. A967 (2017) 720--723, 
\newblock \href {http://arxiv.org/abs/1607.02493} {\path{arXiv:1607.02493}}.

\bibitem{DElia:2016jqh}
M.~D'Elia, G.~Gagliardi, F.~Sanfilippo, Phys. Rev. D95~(9) (2017) 094503,
\newblock \href {http://arxiv.org/abs/1611.08285} {\path{arXiv:1611.08285}}.

\bibitem{Borsanyi:2018grb}
S.~Borsanyi, Z.~Fodor, J.~N. Guenther, S.~K. Katz, K.~K. Szabo, A.~Pasztor,
  I.~Portillo, C.~Ratti, \href {http://arxiv.org/abs/1805.04445}
  {\path{arXiv:1805.04445}}.

\bibitem{Bazavov:2012vg}
A.~Bazavov, et~al., Phys. Rev. Lett. 109 (2012) 192302,
\newblock \href {http://arxiv.org/abs/1208.1220} {\path{arXiv:1208.1220}}.

\bibitem{Bazavov:2015zja}
A.~Bazavov, et~al., Phys. Rev. D93~(1) (2016) 014512,
\newblock \href {http://arxiv.org/abs/1509.05786} {\path{arXiv:1509.05786}}.

\bibitem{Borsanyi:2013hza}
S.~Borsanyi, Z.~Fodor, S.~D. Katz, S.~Krieg, C.~Ratti, K.~K. Szabo, Phys. Rev.
  Lett. 111 (2013) 062005,
\newblock \href {http://arxiv.org/abs/1305.5161} {\path{arXiv:1305.5161}}.

\bibitem{Almasi:2018lok}
G.~A. Almasi, B.~Friman, K.~Morita, P.~M. Lo, K.~Redlich, \href
  {http://arxiv.org/abs/1805.04441} {\path{arXiv:1805.04441}}.

\bibitem{Liu:2016ysz}
S.~Y.~F. Liu, R.~Rapp, \href {http://arxiv.org/abs/1612.09138}
  {\path{arXiv:1612.09138}}.

\bibitem{Liu:2017qah}
S.~Y.~F. Liu, R.~Rapp, Phys. Rev. C97~(3) (2018) 034918,
\newblock \href {http://arxiv.org/abs/1711.03282} {\path{arXiv:1711.03282}}.

\bibitem{Vovchenko:2017gkg}
V.~Vovchenko, J.~Steinheimer, O.~Philipsen, H.~Stoecker, Phys. Rev. D97~(11)
  (2018) 114030,
\newblock \href {http://arxiv.org/abs/1711.01261} {\path{arXiv:1711.01261}}.

\bibitem{Huovinen:2017ogf}
P.~Huovinen, P.~Petreczky, Phys. Lett. B777 (2018) 125--130,
\newblock \href {http://arxiv.org/abs/1708.00879} {\path{arXiv:1708.00879}}.

\bibitem{Parotto:2018pwx}
P.~Parotto, M.~Bluhm, D.~Mroczek, M.~Nahrgang, J.~Noronha-Hostler,
  K.~Rajagopal, C.~Ratti, T.~Schaefer, M.~Stephanov, \href
  {http://arxiv.org/abs/1805.05249} {\path{arXiv:1805.05249}}.

\bibitem{Allton:2003vx}
C.~R. Allton, S.~Ejiri, S.~J. Hands, O.~Kaczmarek, F.~Karsch, E.~Laermann,
  C.~Schmidt, Phys. Rev. D68 (2003) 014507,
\newblock \href {http://arxiv.org/abs/hep-lat/0305007}
  {\path{arXiv:hep-lat/0305007}}.

\bibitem{Gavai:2008zr}
R.~V. Gavai, S.~Gupta, Phys. Rev. D78 (2008) 114503,
\newblock \href {http://arxiv.org/abs/0806.2233} {\path{arXiv:0806.2233}}.

\bibitem{Datta:2016ukp}
S.~Datta, R.~V. Gavai, S.~Gupta, Phys. Rev. D95~(5) (2017) 054512,
\newblock \href {http://arxiv.org/abs/1612.06673} {\path{arXiv:1612.06673}}.

\bibitem{Fodor:2018wul}
Z.~Fodor, M.~Giordano, J.~N. Guenther, K.~Kapas, S.~D. Katz, A.~Pasztor,
  I.~Portillo, C.~Ratti, D.~Sexty, K.~K. Szabo,
\newblock \href {http://arxiv.org/abs/1807.09862} {\path{arXiv:1807.09862}}.

\bibitem{Vovchenko:2018zgt}
V.~Vovchenko, J.~Steinheimer, O.~Philipsen, A.~Pasztor, Z.~Fodor, S.~D. Katz,
  H.~Stoecker,
\newblock \href {http://arxiv.org/abs/1807.06472} {\path{arXiv:1807.06472}}.

\bibitem{Parotto:2018gic}
P.~Parotto,
\newblock \href {http://arxiv.org/abs/1808.03695} {\path{arXiv:1808.03695}}.

\bibitem{Bonati:2013lca}
C.~Bonati, M.~D'Elia, M.~Mariti, F.~Negro, F.~Sanfilippo, Phys. Rev. Lett. 111
  (2013) 182001,
\newblock \href {http://arxiv.org/abs/1307.8063} {\path{arXiv:1307.8063}}.

\bibitem{Levkova:2013qda}
L.~Levkova, C.~DeTar, Phys. Rev. Lett. 112~(1) (2014) 012002,
\newblock \href {http://arxiv.org/abs/1309.1142} {\path{arXiv:1309.1142}}.

\bibitem{Bonati:2013vba}
C.~Bonati, M.~D'Elia, M.~Mariti, F.~Negro, F.~Sanfilippo, Phys. Rev. D89~(5)
  (2014) 054506,
\newblock \href {http://arxiv.org/abs/1310.8656} {\path{arXiv:1310.8656}}.

\bibitem{Bali:2013owa}
G.~S. Bali, F.~Bruckmann, G.~Endrodi, A.~Schafer, Phys. Rev. Lett. 112 (2014)
  042301,
\newblock \href {http://arxiv.org/abs/1311.2559} {\path{arXiv:1311.2559}}.

\bibitem{Bali:2014kia}
G.~S. Bali, F.~Bruckmann, G.~Endrodi, S.~D. Katz, A.~Schaefer, JHEP 08 (2014)
  177,
\newblock \href {http://arxiv.org/abs/1406.0269} {\path{arXiv:1406.0269}}.

\bibitem{Bali:2011qj}
G.~S. Bali, F.~Bruckmann, G.~Endrodi, Z.~Fodor, S.~D. Katz, S.~Krieg,
  A.~Schafer, K.~K. Szabo, JHEP 02 (2012) 044,
\newblock \href {http://arxiv.org/abs/1111.4956} {\path{arXiv:1111.4956}}.

\bibitem{DElia:2018xwo}
M.~D'Elia, F.~Manigrasso, F.~Negro, F.~Sanfilippo, Phys. Rev. D98~(5) (2018)
  054509,
\newblock \href {http://arxiv.org/abs/1808.07008} {\path{arXiv:1808.07008}}.

\bibitem{Yamamoto:2013zwa}
A.~Yamamoto, Y.~Hirono, Phys. Rev. Lett. 111 (2013) 081601,
\newblock \href {http://arxiv.org/abs/1303.6292} {\path{arXiv:1303.6292}}.

\end{thebibliography}





\end{document}